# Controlled Doping of Double Walled Carbon Nanotubes and Conducting Polymers in a Composite: An in situ Raman Spectroelectrochemical Study.

**Martin Kalbáč*[a,b], Ladislav Kavan,[a] and Lothar Dunsch[b]**

[a] *J. Heyrovský Institute of Physical Chemistry, v.v.i. Academy of Sciences of the Czech Republic, Dolejškova 3, CZ-18223 Prague 8, Czech Republic. Tel: 420 2 6605 3804; Fax: 420 2 8658 2307; E-mail: kalbac@jh-inst.cas.cz*

[b] *Leibniz Institute of Solid State and Materials Research, Helmholtzstr. 20, D - 01069 Dresden, Germany. Fax: 49 351 4659 811 ; Tel: 49 351 4659 660; E-mail: l.dunsch@ifw-dresden.de*

## Abstract

The interaction of double wall carbon nanotubes (DWCNTs) and the conducting polymer poly(3,4-ethylenedioxythiphene/polystyrenesulfonate (PEDOT/PSS) was studied by in-situ Raman spectroelectrochemistry. The mixing of DWCNTs with PEDOT/PSS caused a partial doping of the outer tube of DWCNTs, which was indicated by the relative change of the Raman intensity of the DWCNTs' features. On the other hand, the bands corresponding to inner tubes of DWCNTs and to the polymer were almost untouched by assembling both species into a composite. The in situ Raman spectroelectrochemical experiments have shown that the changes in electronic structure of inner tubes of DWCNTs embedded in PEDOT/PSS matrix are dependent on the doping level. While at the low doping level of the composite, the Raman features of inner tubes of DWCNTs do not change significantly, at high doping level they reflect the changes caused by the applied electrochemical potential similar to that observed in the polymer-free DWCNTs.

## Introduction

An important application of carbon nanotubes consists in their use for the fabrication of composites with polymers. The embedding of single wall carbon nanotubes (SWCNTs) into a conducting polymer leads to the change of electronic structure of SWCNTs and their mechanical properties. These electronic effects should be suppressed for inner tubes in the case of composites with double wall carbon nanotubes (DWCNTs). Indeed, the first studies on nanocomposites consisting of DWCNTs and an epoxy matrix exhibited considerable improvements of the properties of the resulting composite material [1]. The high quality DWCNTs can be prepared by a vacuum pyrolysis of fullerene peapods[2] or by chemical vapor deposition (CVD)[3]. The CVD production of DWCNTs is obviously cheaper than the first procedure and can be scaled up for industrial purposes. However, depending on experimental conditions used in the CVD process, a mixture of carbon nanostructures is obtained containing SWCNTs, DWCNTs and multiwalled carbon nanotubes (MWCNTs). Furthermore, the CVD synthesis of nanotubes provides samples of a relatively wide diameter distribution and it is not always easy to distinguish by Raman spectroscopy the DWCNT's inner/outer tubes and the SWCNTs, respectively [4]. This complicates the interpretation of Raman spectra. Therefore the use of ex-peapod DWCNTs is more convenient for fundamental studies.

Poly(3,4-ethylenedioxythiophene) (PEDOT) is a stable polymer, which exhibits relatively high electrical conductivity [5]. It is stable even during electrochemical charging and discharging [6], which allows extensive studies of the charged states of this polymer. Doped PEDOT films are optically well transparent, which makes them suitable for applications in electrochromic devices, solar cells, antistatic films or light emitting diodes. Furthermore, it is well known that the doping of the polymer or carbon nanostructure has a great impact on the electronic properties of nanoscaled materials. Thus the ability to control the doping level of a composite is critical for applications in nanoelectronics. In the case of nanocomposites the doping also plays a very important role for the state of the components. Since there is a charge transfer between the polymer and the carbon nanotubes in a composite, the external doping has an impact on the electronic structure of both components [7-10].

The in situ spectroelectrochemistry is a method of choice for studying PEDOT films [11-14] and carbon nanostructures [15-20]. We have demonstrated previously, that the in situ Raman spectroelectrochemistry is a precise and efficient method for the control and characterization of the doping level of the polymer/SWCNTs composites [10]. Hence, it was applied previously to the studies of the composites of single walled carbon nanotubes with PEDOT/PSS [10]. We showed that a simple mixing of pristine SWCNTs with PEDOT/PSS caused already a partial doping of SWCNTs. Furthermore, it was shown that the electronic structure of the embedded SWCNTs develops in different way during electrochemical doping than in the case of pristine SWCNTs. This was indicated by the change of the relative intensity of the Raman features of SWCNTs.

The inner tubes in DWCNTs are shielded by the outer tubes [15,16,21]. This has important consequences for the variation of Raman spectra during electrochemical charging. The shielding of inner tubes is also an argument for preferable use of DWCNTs in nanotube composites, since the interaction between the polymer and the outer tube would not presumably disturb the electronic structure of inner tube. Thus it is challenging to investigate the embedding of DWCNTs into a polymer matrix. In this work we present the first in situ spectroelectrochemical study of double walled carbon nanotubes in polymer composites. Our results show that the inner tubes of DWCNTs, indeed, do not change markedly their electronic structure if DWCNTs are embedded in a PEDOT/PSS matrix. However we demonstrate by the in-situ Raman spectroelectrochemistry that this situation may change if a suitable electrochemical charging is applied.

**Experimental section**

The ex-peapod DWCNTs samples were obtained by heat treatment of peapods at 1200°C for 8 h in vacuum ($10^{-6}$ Pa) [2]. The fullerene peapods were available from our previous work [22]. A thin-film of DWCNT was prepared by evaporation of the sonicated (approx. 15 min.) ethanolic slurry of the DWCNTs on a Pt electrode in air. A thin film electrode of PEDOT/PSS was prepared in a similar way by evaporation of an aqueous solution of PEDOT/PSS (Baytron, Bayer AG) on the Pt electrode. For the preparation of composites the pure DWCNTs film on the Pt electrode was soaked with the aqueous solution of PEDOT/PSS. Afterwards the composites have been dried (these samples were further referred to as DWCNTs-PEDOT/PSS composites). The film electrodes were outgased overnight at 90°C in vacuum and then mounted in a Raman spectroelectrochemical cell in a glove box. The cell was equipped with a Pt-counterelectrode and an Ag-wire pseudo-reference electrode. 0.2 M $LiClO_4$ in dry acetonitrile was used as the supporting electrolyte solution. Electrochemical experiments were carried out using the PG 300 (HEKA) or the EG&G PAR 273A potentiostats.

The Raman spectra were measured on a T-64000 spectrometer (Instruments SA) interfaced to an Olympus BH2 microscope (the laser power impinging on the sample or cell window was between 1-5 mW). Spectra were excited by a $Kr^+$ laser at 1.92 eV (Innova 305, Coherent). The Raman spectrometer was calibrated before each set of measurements by using the $F_{1g}$ line of Si at 520.2 $cm^{-1}$.

**Results and discussion**

The interaction of DWCNTs and PEDOT/PSS has been studied in a composite made by simple soaking of solid DWCNTs film with aqueous solution of PEDOT/PSS (see Experimental Section for details). This method has been demonstrated to lead to a penetration of the polymer into the SWCNTs buckypaper or thin film [10,23]. Figure 1 shows the Raman spectra of pristine DWCNTs, PEDOT/PSS and the composite DWCNTs-PEDOT/PSS, which were recorded using 1.92 eV laser excitation energy. The spectrum of DWCNTs exhibits typical features of the radial breathing mode (RBM), D, tangential mode (TG) and G' modes. The RBM frequency of isolated (non-bundled) SWCNT is inversely proportional to the tube diameter:

$$\varpi = \frac{C}{d} \tag{1}$$

where the constant C was reported to be in the range of 224 to 251 nm.cm$^{-1}$ (Refs. [24-26]). However, the highest values (around 250 nm.cm$^{-1}$) are no longer accepted [20]. The DWCNTs have been prepared from peapods with a known diameter distribution. The simple comparison of the Raman spectra of the parent peapods (not shown) and the DWCNTs at the particular laser excitation energy provides enough information, for the RBM assignment to the outer and inner tubes, respectively. Obviously, the diameter of the inner/outer tubes is determined by the dimensions of the parent peapod. The optimum diameter of the SWCNT precursor for the formation of $C_{60}$@SWCNT peapod is around 1.4 nm [27]. Thus, the optimum diameter of inner tubes is similar to the diameter of $C_{60}$, which is ca. 0.7 nm. Hence, the RBM modes are well distinguished for inner and outer tubes. At a laser excitation energy of 1.92 eV (Figure 1) the outer tubes' RBM are found in the region between 150-200 cm$^{-1}$ while the RBM of the inner tubes are in the region between 280-370 cm$^{-1}$. Since the positions of the RBM signals of inner/outer tubes in ex-peapod DWCNTs are well separated, it is easy to study the electronic structure of inner and outer tubes and its variation separately. The diameter of the outer tubes of DWCNTs was calculated to be about 1.5 nm using Eq. 1. According to the Kataura plot[28], the optical transitions in 1.5 nm-metallic tubes resonate with the 1.92 eV laser energy. Indeed the TG mode exhibits the Breit-Wigner-Fano (BWF) broadening typical for metallic tubes. On the other hand, in the case of inner tubes, only semiconducting tubes are in resonance with the 1.92 eV laser energy. However, the TG mode of inner tubes is overlapped by the TG mode of outer tubes and thus it is impossible to perform its detailed analysis in the undoped state.

The main Raman features of PEDOT/PSS were assigned as follows [11]: 558, 989 cm$^{-1}$ (oxyethylene ring deformation), 699 cm$^{-1}$ (symmetric C-S-C deformation), 1100 and 1130 cm$^{-1}$ (C-O-C deformation), 1200-1280 cm$^{-1}$ (C$_\alpha$-C$_{\alpha'}$ stretching + C-H bending), 1365 cm$^{-1}$ (C$_\beta$-C$_{\beta'}$ stretching), 1435 cm$^{-1}$ (symmetric C=C) and 1510-1580 cm$^{-1}$ (antisymmetric C=C stretching).

The Raman spectra of the DWCNTs-PEDOT/PSS composite correspond roughly to the superposition of the spectra of DWCNTs and PEDOT/PSS. However, a more detailed analysis shows some slight differences caused by an interaction between the polymer and the outer tubes of DWCNTs. This is similar to the case of SWCNTs/polymer composites [10]. For example, there is a small shifted upward (by 2 cm$^{-1}$) of the outer tube nanotube's RBM both for SWCNTs[10] and DWCNTs after their embedding into the PEDOT/PSS polymer matrix. This confirms a good mixing of DWCNTs with PEDOT/PSS polymer. On the other hand, there is no significant shift of the RBM bands of inner tubes. This is an indication that (without external doping) the inner tube does not feel the interaction of the outer tube with polymer.

There are also changes in the TG mode of the DWCNT upon embedding into the PEDOT/PSS polymer matrix. The changes are expected to be dominated by the outer tube features. However, it is difficult to evaluate these results since the TG modes of inner and outer tube overlap in the virgin state of the composite.

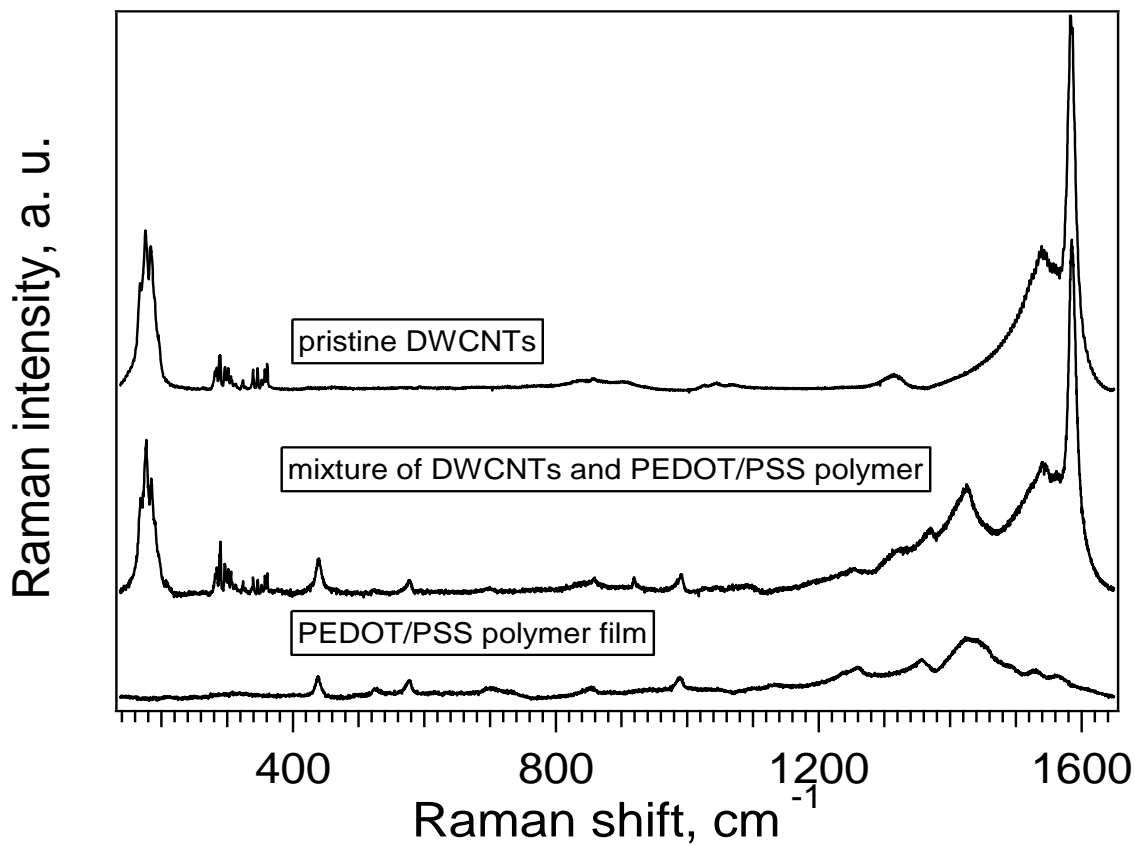

**FIGURE 1.** Raman spectra (excited at 1.92 eV) of pristine DWCNTs, mixture of DWCNTs and PEDOT/PSS polymer and PEDOT/PSS polymer film (from top to bottom) on a platinum electrode surface. Intensities of spectra were normalized using the $F_{1g}$ line of Si at 520.2 cm$^{-1}$. Spectra are offset for clarity, but the intensity scale is identical for all spectra.

**In-situ Raman spectroelectrochemistry**

Figure 2 shows the in-situ Raman spectroelectrochemical data of a DWCNTs-PEDOT/PSS nanocomposite. Due to the limited electrochemical stability of PEDOT/PSS as compared to the nanotube the spectroelectrochemical experiments were performed between +1.4 and –1.0 V (vs. Ag pseudoreference electrode). The measurements started in each case at the applied electrode potential close to the open circuit potential to minimize the change of the electronic structure and the doping level of the components at the beginning of the experiment. The potential was increased in 0.2 V steps toward +1.4 V and then decreased to –1.0 V. At each step the Raman spectra were recorded, while keeping the potential constant. No significant current flow has been detected during the measurement of Raman spectra, which ensures that the spectra were recorded at the equilibrium state.

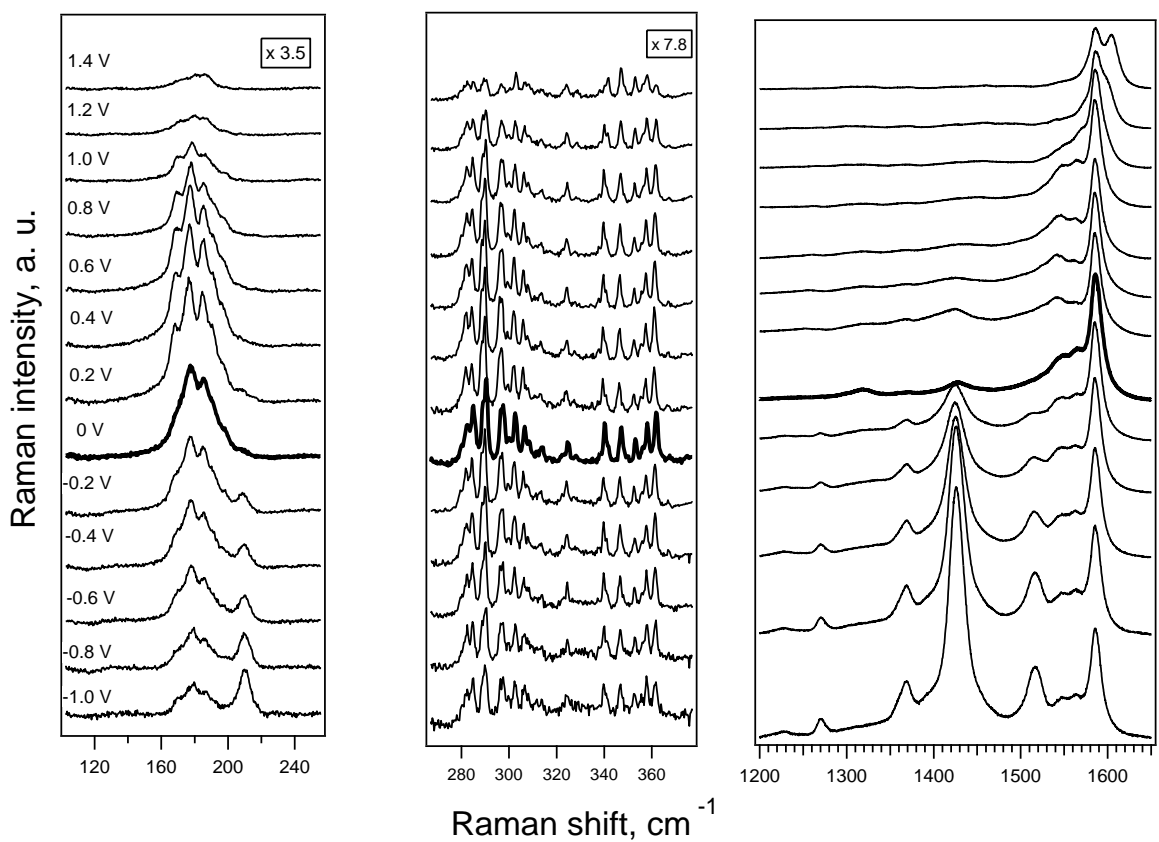

**FIGURE 2.** Potential dependent Raman spectra (excited at 1.92 eV) of DWCNTs-PEDOT/PSS on a Pt electrode in 0.2 M $LiClO_4$ + acetonitrile solution. The electrode potential varied by steps of 0.2 V from 1.4 to -1.0 V vs. Ag pseudoreference electrode for curves from top to bottom. Spectra are offset for clarity, but the intensity scale is identical for all spectra in the respective window.

There is a dramatic increase in the intensity of the polymer bands during cathodic doping.(Figure 2) This effect is associated with the resonant Raman effect [14]. The electrochemical reduction of the polymer leads to a strong increase of the broad optical absorption band centered at the wavelength of 600 nm [14]. Thus the Raman spectra excited near this wavelength become resonantly enhanced. These results are in agreement with those previously measured on SWCNTs-PEDOT/PSS composites [10]. This confirms that the interactions between the outer tubes of DWCNTs and the polymer are not significantly perturbed by the presence of inner tubes.

Another significant effect of electrochemical doping is a subsequent bleaching of all main features of the outer tubes of DWCNTs (the RBM, D and TG modes). The charging-induced bleaching of the modes of the DWCNTs' outer tubes matches the previously published results on SWCNTs and the polymer-free DWCNTs [17,29]. As the electrochemical doping leads to a shift of the Fermi level, the van Hove singularities of the nanotubes are filled/depleted upon cathodic/anodic charging. This erases the Raman resonance enhancement due to the loss of optical absorptions between the van Hove singularities, and finally it leads to the subsequent bleaching of all features of outer tubes of DWCNTs in the Raman spectra. Furthermore, there are shifts in the Raman band positions during the electrochemical doping. The most obvious effect is the upshift of the TG mode of the outer tubes of DWCNTs at high anodic potentials. This confirms that the DWCNTs are in good electrical contact with the Pt electrode even if they are assembled with the polymer.

In pristine DWCNTs, the inner tube modes of DWCNTs are extinguished less than the bands of the outer tubes when applying an electrochemical potential, because the inner tubes are shielded by the outer tubes [16,21]. This effect is most obvious on the inner-tube's RBM, but it can be also followed for the D, TG, G' modes [15] and even for the intermediate frequency modes [30]. The D, TG, and G' lines shift upward at high positive potentials. This is a consequence of the stiffening of the CC bonds due to the electrochemical doping. However, the latter effect applies only for the outer tubes and not for the inner tubes. Therefore each of the D, TG, G'bands of DWCNTs splits into a doublet at high anodic potentials [15]. This is indeed observed even for DWCNTs embedded in the PEDOT/PSS polymer matrix. There is a clear split of the TG mode in Raman spectrum of the composite at +1.4 V (Figure 2). Nevertheless the shift of the TG mode is observed only at high doping levels. Thus it is not surprising that for the inner tubes the TG mode does not change its frequency. On the other hand the intensity of the RBM bands is more sensitive, which is traceable already at the low level doping. The intensities of the inner tube bands of DWCNTs-PEDOT/PSS change as a result of stronger electrochemical doping. This shows that the electronic structure of both the outer and the inner tubes of DWCNTs is influenced by electrochemical charging even if DWCNTs are embedded into a nanocomposite.

Figure 3 gives a detailed comparison of the spectra of pristine DWCNTs and DWCNTs embedded in the PEDOT/PSS polymer matrix at the open circuit potential (OCP), 0.6 V and 1.2 V (vs. Ag pseudoreference electrode).

At the potential of 0.6 V the spectra of the pristine DWCNTs exhibit a higher level of doping than the spectra of DWCNTs embedded in the PEDOT/PSS polymer matrix. This is indicated by a smaller intensity of the RBM bands and a difference in the shape of the TG mode. The intensity of the RBM bands of inner tubes of DWCNTs embedded in the PEDOT/PSS polymer matrix do not even change significantly at the potential of 0.6 V with respect to the spectra measured at OCP. This indicates that, at this doping level, the inner tube does not reflect the changes of the electronic state of the outer tube. However, the situation at a potential of 1.2 V is different. The spectra of DWCNTs and DWCNTs embedded in the PEDOT/PSS polymer matrix are almost identical. In other words at a high doping level the change of electronic structure of the outer tubes is reflected by the inner tubes independently whether DWCNTs are embedded in the polymer or not.

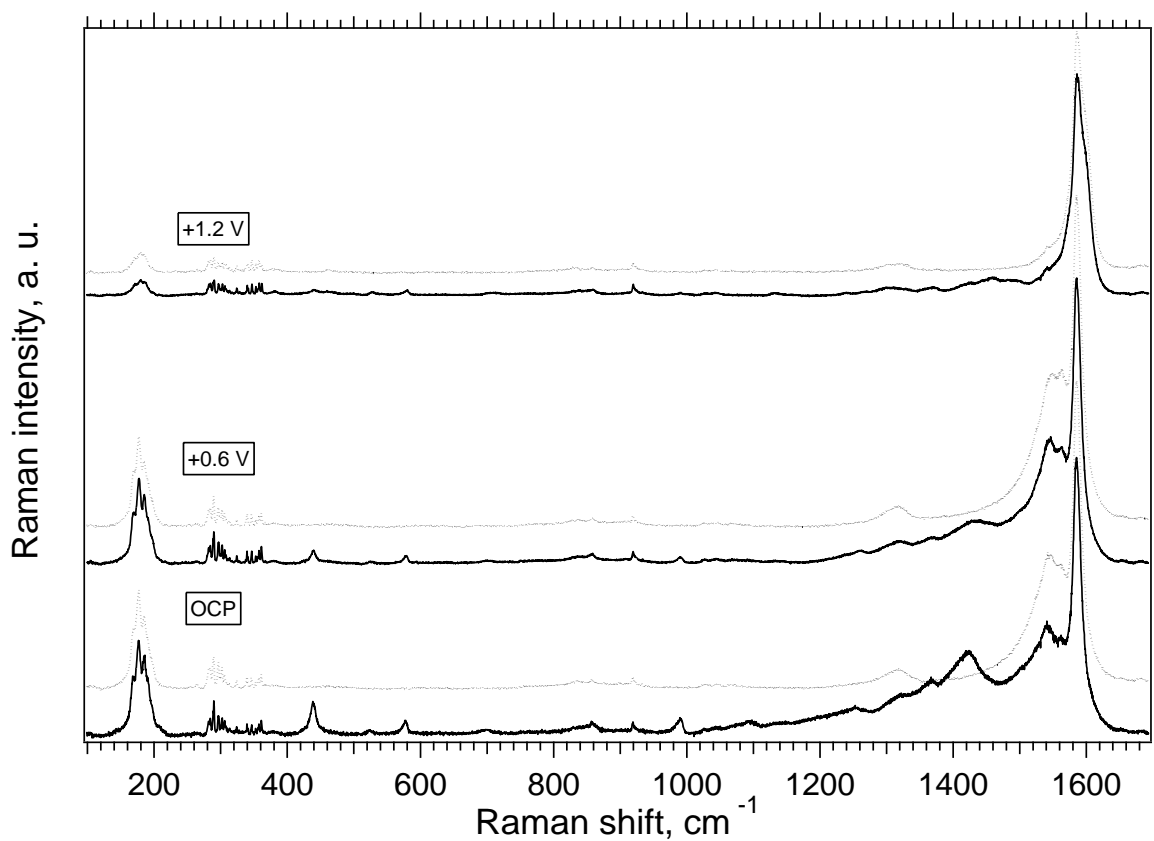

**FIGURE 3.** Potential dependent Raman spectra excited at 1.92 eV of DWCNTs (dotted line) DWCNTs-PEDOT/PSS (solid line) on a Pt electrode in 0.2 M $LiClO_4$ + acetonitrile solutions. The electrode potentials are +1.2 V, +0.6 V vs. Ag and OCP (open circuit potential) for curves from top to bottom. Spectra are offset for clarity, but the intensity scale is identical for all spectra. Intensities of spectra were normalized using the $F_{1g}$ line of Si at 520.2 cm$^{-1}$.

**Conclusions**

We have studied DWCNTs- PEDOT/PSS nanocomposites using Raman spectroscopy and in situ Raman spectroelectrochemistry. The mixing of DWCNTs with polymer PEDOT/PSS leads to changes in the Raman spectra of DWCNTs. The bands of the polymer are almost intact. The behavior of DWCNTs- PEDOT/PSS nanocomposite was studied with respect to the behavior of its components under electrochemical charging. The Raman lines attributed to outer tubes of DWCNTs in the DWCNTs-PEDOT/PSS composite exhibited a similar behavior upon charging as it was found in the case of the pristine DWCNTs. The lines of outer tubes are subsequently bleached as far as an electrochemical potential is applied. Nevertheless there is a slight delay in the bleaching of the bands in the case of the nanocomposites compared to those of pristine DWCNTs. At this level of doping the features of inner tubes of DWCNTs embedded in composite do not change significantly.

However the situation is different at higher electrode potentials. There is no difference in the spectra of pristine DWCNTs and in the DWCNTs embedded in the DWCNTs-PEDOT/PSS composite. Thus the behavior of the nanocomposite during electrochemical charging is dominated by the doping of both components of the DWCNT, which induces changes in electronic structure of individual components of nanocomposite.

**Acknowledgement**

This work was supported by the Deutsche Forschungsgemeinschaft, the Academy of Sciences of the Czech Republic (contracts KJB400400601, IAA400400804 and KAN200100801), the DFG-GACR project (contract No 203/07/J067) and by the Czech Ministry of Education, Youth and sports (contract No. LC-510). M. Kalbac acknowledges a grant of the Alexander von Humboldt Foundation and the support of the IFW.